# Ultrafast observation of electron hybridization and in-gap states formation in Kondo insulator $SmB_6$

Sanjay Adhikari[1], Yanjun Ma[2], Zachary Fisk[3], Jing Xia[3], Chang-Beom Eom[2], Cheng Cen[1*]

[1]*Department of Physics and Astronomy, West Virginia University, Morgantown, West Virginia 26506, USA*

[2]*Department of Material Science and Engineering, University of Wisconsin-Madison, Madison, Wisconsin 53706, USA*

[3]*Department of Physics and Astronomy, University of California, Irvine, California 92697-4575, USA*

*Correspondence to chcen@mail.wvu.edu*

**Abstract**

$SmB_6$ is a promising candidate for topological Kondo insulator. In this letter, we report ultrafast carrier dynamics of $SmB_6$. Two characteristic temperatures: $T_1$=100 K and $T_2$= 20 K are observed. $T_1$ corresponds to the opening of the *f-d* hybridization gap revealed by an abrupt disappearance of terahertz *f*-band plasmon oscillations. Between $T_1$ and $T_2$, a "phonon bottleneck" effect dominates the photocarrier relaxation processes. Below $T_2$, we observe the formation of in-gap states, which are strongly affected by optically injected hot electrons and the transient electron temperature change.

**Introduction**

Kondo insulators (KI), in particular $SmB_6$, are compounds in which strong correlation effects give rise to an insulating ground state at low temperatures. In general, the strong interaction within the localized periodic dense array of *f*-magnetic-moments (so-called Kondo lattice) leads to reconstructions of the electronic structure and opens up [1] an energy gap at low temperatures due to the hybridization between conduction electrons and the highly renormalized *f*-electrons. When the Fermi level lies within the gap, a KI is formed. Recent theories of topological Kondo insulator (TKI) [2-12] have predicted, in $SmB_6$, the existence of a topological nontrivial surface state with odd number of Dirac surface bands. The hybridization and odd parity wavefunction lead to strong spin-orbit coupling and give rise to a topological surface state that dominates electron conduction at low temperatures. $SmB_6$ is also predicted to be a strong topological insulator (TI) [3,4]. First principle calculations [4-7,9-11] have proposed a surface state with three Dirac bands at Γ and X/Y points in the (100) surface.

In the past few years, tremendous experimental progress has been made regarding $SmB_6$. Well-designed transport experiments revealed a low temperature resistance plateau associated with the surface layer [13-17]. Point-contact spectroscopy [18], scanning tunneling spectroscopy (STS) [19,20], angle resolved photoemission spectroscopy (ARPES) [21-26] and magnetometry [27] provided further evidences of the surface state. Sensitivity to magnetic dopants, Dirac-like dispersion and spin helicity have been discovered [16,21,22,27] in support of the TKI theory. In the meanwhile, controversies still exist. Nontopological surface states associated with the Boron dangling bond, common in hexaborides [28,29], bring additional complexities to the surface states interpretation. Evidences of surface Boron suboxide formation [30], chemical potential shift at polar and non-polar surfaces [31] and polarity driven surface metallicity [32] were found in $SmB_6$. Metallic bulk states [33] and impurity states based hopping mechanism[34] were also suggested as alternative contributors to the low temperature resistance plateau. The entanglement of surface state topology with strongly correlated bulk states as well as the complicated surface chemistry makes $SmB_6$ a highly attractive system for more in depth investigations.

Most of the measurements mentioned above address equilibrium states, where material properties are often described by effective thermodynamic parameters. On the contrary, pump-probe technique abruptly disturbs the material equilibrium by laser pulses and studies the following ultrafast re-equilibrium process as a function of time. By examining various characteristic dynamics in time domain, this technique can help distinguish many intrinsic degrees of freedom that are entangled at equilibrium. Examples include the study of electron-phonon interaction [35-38], order parameter fluctuation [39], pseudogap formation [40], and spin-orbit coupling [41] in many strongly correlated materials. In $SmB_6$, besides the pioneering work by Demsar. *et. al.* [36], recent works of time resolved Terahertz spectroscopy [42-44] and pump-probe photoemission spectroscopy [45] have also provided valuable insights into the low temperature band structure and the surface band bending. In this letter, photoexcited electron dynamics in $SmB_6$ are investigated by optical pump-probe technique. Observations of the characteristic *f*-electrons plasmon help pin-point the hybridization gap opening temperature. Probing the photocarrier relaxation in time domain allows us to quantify the gap and sensitively detect the emergence of in-gap states at much lower temperature. Varying both the equilibrium temperature as well as the non-thermal distribution of electrons and phonons, an electron correlation based origin of the in-gap states is revealed. To evaluate the robustness of the observed properties, experiments are performed on two types of samples with distinct crystalline characters: single crystal and polycrystalline (as grown) thin film.

**Experimental Setup**

SmB$_6$ single crystals were grown using the aluminum flux method similar to previous works [14,16,17]. Thin films of SmB$_6$ are deposited on top of MgO substrates by pulsed laser deposition (PLD) [46]. Resistances in both single crystal and thin film samples saturate below 4 K (Fig. S1) [46], indicating an insulating bulk and the possible formation of surface states at low temperatures[14,16]. Our pump probe setup consists of 35 fs probe pulses centered at 840 nm with a 74 MHz repetition rate, and pump pulses from their second harmonic (420 nm). Both pump and probe are focused to the sample surface with a single aspheric lens. Sample temperature is varied between 5 K and 294 K in a close cycle cryostat with pressure maintained at 10$^{-5}$ mbar level. All the data we present here are robust against full range thermal cycles.

**Data and Discussion**

Time-resolved measurements of pump induced (PI) probe reflectivity changes ($\Delta R/R$) at different temperatures are displayed in Figure 1. At zero delay, $\Delta R/R$ rises sharply due to the hot electrons generation. Following the initial excitation, electrons quickly thermalize through electron-electron (*e-e*) scattering and phonon emissions. During this process, energies of hot electrons experience significant relaxation and the photocarrier density is greatly amplified, which is represented by the sub-picosecond rise of the pump-probe signal close to zero delay. During the slow recovery process, both monotonic decays and coherent oscillations are observed.

Between 300 K and 120 K, a 1.7 THz oscillation signal $S_1$ is observed in the single crystal (Fig.2 (a,b)). $S_1$ is temperature independent above 120 K, but exhibits abrupt decoherence and disappears completely below $T_1$=100K. Frequency of $S_1$ remains constant when varying the probe wavelength and the incidence angle, which indicates that $S_1$ corresponds to a mode almost non-dispersive at the zone center. Optical phonons fit such description. However, 1.7 THz is substantially lower than all the known optical phonon branches [47-49]. The strong temperature dependence also rules out defect related modes. Instead, we attribute $S_1$ to the bulk plasmon resonance of the strongly screened 4*f* heavy fermions (Fig.2(c)). 4*f* electron plasmon frequency can be expressed as:

$$\omega_p = \sqrt{n_{4f} e^2 / m^* \epsilon_{op} \epsilon_0} \quad (1)$$

Its frequency is determined by electron density ($n_{4f}$), effective mass ($m^*$), and the effective permittivity relating to the screening by *f-d* interband transition ($\epsilon_{op}$). A similar frequency of this resonance (5.3 meV = 1.3 THz) is reported by spectroscopy measurements [50]. We note that, $n_{4f}$ and $\epsilon_{op}$ both sensitively depends on the samples' chemical potentials, which may vary due to differences in synthesis conditions or surface adsorptions in ambient environments. It is very likely that the small frequency difference between our measurement and the earlier spectroscopy report is associated with such chemical potential variation. The constant frequency of $S_1$ above 120 K indicates a metal-like band structure. The drastic decoherence of $S_1$ signal around $T_1$ marks the opening of the *f-d* hybridization gap (Fig.2(c)). As temperature decreases, the heavy *f* electrons strongly hybridize with the light *d* electrons and opens a gap. Characteristic *f*-band plasmon thus disappears, and new electron bands form with mixed *f-d* characters. Similar signal is observed in the thin film but with a different frequency of 2.8 THz. Defects present in the thin film can raise the chemical potential and lead to a larger electron density. In addition, the close proximity with wide bandgap MgO substrate is expected to further reduce the screening strength (smaller $\epsilon_{op}$). Both effects are expected to give rise to a higher plasmon frequency.

While coherent oscillation signals reveal the gap opening at $T_1$=100 K, monotonic decays in Δ*R/R* at low temperatures allow further characterizations of the hybridization gap. We analyze the data using a phenomenological Rathwarf-Taylor (RT) model [51], which were developed for systems with a narrow bandgap [36,52-54]. In these systems, zero density of states inside the gap forbids the excited electron-hole pairs from releasing energy gradually through sequential emissions of low energy phonons. While the emission of phonons with energy above the bandgap is allowed, these high energy phonons can in return re-excite new electron-hole pairs. Therefore the complete relaxation of photocarriers cannot occur before the fully decay of high energy phonons The later takes place mainly through spatial diffusion or scattering with low energy phonons. Such mechanism, often referred to as "phonon bottleneck", is often present in systems with discontinuous or discrete energy distributions [55].

At low pump fluence, maximum intensity ($I_{max}$) of Δ*R/R* signal represents the photocarrier density after fast *e-e* thermalization ($n_{pc}$), which depends strongly on the thermal electron distribution prior to the excitation. Described by the RT model, the temperature dependence of the thermal carrier density can be extracted from: $n(T) \propto I_{max}(T \to 0)/I_{max}(T)$ [36,51]. At low temperatures, $I_{max}$ increases dramatically (Fig.1), consistent with the reduction of *n* in presence of a hybridization gap $E_g$ [36]:

$$n(T) \propto T^{1/2} \exp(-\frac{E_g(T)}{2kT}) \qquad (2)$$

Assuming a gradually changing gap energy, which is supported by STS measurements [19,20], we take a linear approximation of $E_g(T) \approx E_g(20) - c(T-20)$. Equation (2) becomes

$$n(T) \propto T^{1/2} \exp(-\frac{E_g(20) - c(T-20)}{2kT}) = \exp(\frac{c}{2k})T^{1/2} \exp(-\frac{E_g(20) + 20c}{2kT}) \qquad (3)$$

In both single crystal and thin film samples, $n(T)$ agrees very well with this RT model description between 20 K and 100 K (Fig. 3). Above 100 K, clear deviations from the RT model are observed, which is consistent with the high temperature gap-less metallic state concluded from the plasmon measurements. Least square fitting yields $E_g^* = E_g(20) + 20c$ of 20.8 meV for single crystal and 14.5 meV for thin film. Since gap energy is positive between 20K and 100K, the value of *c* is restricted by $0 \leq c \leq \frac{E_g(20)}{100-20}$. As a result, we have $0.8 E_g^* \leq E_g(20) \leq E_g^*$. Therefore, the gap energy at 20 K can be characterized by $E_g^*$ with a less than 20% error. In the meanwhile, the transport activation energy extracted from electrical measurements (Fig.S2 [46]) are considerably lower (9.2 meV for single crystal and 5.7 meV for thin film). The gap energy obtained from pump-probe and transport measurements interestingly fall into the two distinct categories found in literature reports: 15 meV – 20 meV [22,24,25,56-58], and less than 10 meV [13,14,20,33,58-60]. Seeing values from the two categories simultaneously on the same samples implies that, the discrepancy in gap measurements may come from intrinsic reasons. Several theoretical works have predicted the indirect nature of the hybridization gap [5,6,11]. This can lead to measurement specific biases and explains why photon based experiments, emphasizing on momentum-conserved transitions, tend to yield larger band gap values comparing to the transport measurements.

Temporal profile of $\Delta R/R$ decay can be fitted by a bi-exponential function with two decay constants $\tau_1$ and $\tau_2$ ($\tau_1 \sim 10^0$ ps, $\tau_2 \sim 10^2$ ps). Within the frame of RT model, the fast decay time $\tau_1$ is inversely proportional to $2n+n_{pc}$ [36,53]. As the thermal carrier density $n$ decreases, the fast decay is expected to slow down. The much longer decay time $\tau_2$ is related to a slower relaxation process involving carrier diffusion and lattice cooling. These processes also tend to be suppressed at low temperatures. Therefore, RT model predicts a slowing carrier relaxation as temperature decreases. This is exactly what's observed between 20 K and 100 K (Fig.1). However, below $T_2$=20 K, temporal decay dramatically accelerates in the single crystal (Fig.1 (a)). Similar but less significant trend is also observed in the thin film (Fig. 1(b)). Such features raise the question of whether RT model is still valid below $T_2$.

To answer this question, we investigate the pump-probe signal's fluence dependence. At 20 K, the temporal profile of signal decay is laser fluence independent (Fig. 4(a), inset). This signature is consistent with RT model when photoexcitation can be considered as weak perturbation to the thermal carrier background. In contrast, at 5 K, decay profile becomes highly fluence dependent (Fig. 4(a)). Data taken with $F$ larger than 175 nJ/cm$^2$ have slow decay profiles almost identical to 20 K cases. This is consistent with results reported by Demsar, *et. al* [36] that covers only the higher fluence range. The accelerated decay is only observed below a fluences threshold of $F_c$=175 nJ/cm$^2$.

At very low temperatures, number of thermal carriers practically diminishes. And as a result, the "weak perturbation" assumption is no longer accurate in terms of pulse excitation. In other word, laser "heating" effects need to be included when assessing the validity of RT model. This includes the raise of sample thermal equilibrium temperature ($T^*_{eq}$) and impulsive heating of local electron and lattice temperatures ($T^*_e, T^*_p$). We model $T^*_{eq}$ in our cryostat setup by solving thermal conduction equation using finite element method (Fig. S5 [46]). Below 175 nJ/cm$^2$, we find only a weak heating of $T^*_{eq}$ by less than 2 K. In contrast, transient temperatures $T^*_e$ and $T^*_p$ are more significantly affected by the laser heating. Following the effective temperature method developed by Parker [61], we define $T^*_e$ and $T^*_p$ to characterize the out-of-equilibrium transient distributions of electrons and phonons. Their values at the quasi-equilibrium between electrons and above-bandgap phonons (phonon bottleneck) are plotted in Figure. 4(b). The calculation details are described in supplementary materials [46]. $T^*_e$ raises with increasing $F$, and is $T$-independent in the lowest temperature range where photoelectrons dominate. At lowest temperatures, phonon energy capacity is very low. Due to the phonon bottleneck, high energy phonons emitted bring the phonon population even more out of equilibrium than the photoexcited electrons. As a result, $T^*_p$ takes an interesting upturn at the lowest temperatures. T-independent electron temperature is inconsistent with the $T$-dependent decay profile change (Fig.1). The upturn of phonon temperature appears more relevant. Nevertheless, the heating of $T^*_p$ is more significant at larger

*F*, which contradicts the low fluence requirement for observing the decay acceleration. Therefore, we conclude that, even when heating effects are included, the narrow bandgap RT model is no longer valid at the lowest temperatures.

The drastically accelerated relaxation dynamics below $T_2$, however, can be explained considering the emergence of in-gap states [3,4,6,16,19,22,24]. With nonzero density of states inside the gap, "phonon bottleneck" can be removed. In-gap states provide effective pathways for conduction band electrons to relax their energy through the scattering with low energy phonons (Fig. 4(c)). We propose the following mechanism to understand the low laser fluence required for the in-gap states observation. Excited by the visible pump pulse, $T_e^*$ first rises to a very high value due to hot electrons generation. During this initial high $T_e^*$ period, in-gap states are destroyed and RT model still rules. After *e-e* thermalization and high energy phonon emissions, $T_e^*$ at quasi-equilibrium with high energy phonon is determined by the laser fluence (Fig. 4(b)). At high *F*, $T_e^*$ at quasi-equilibrium is higher than $T_2$, and therefore prohibits the formation of in-gap states at less than 100 ps time scale. In contrast, when *F* is low, $T_e^*$ quickly settles to a value lower than $T_2$, which allows the development of in-gap states and the more efficient energy transfer to the lattice (Fig. 4(c)). According to this model, the calculated fluence threshold corresponding to a critical $T_e^*$ of 20 K is around $F_c$=175 nJ/cm$^2$, matching perfectly with the experimental observation in single crystal. In the thin film sample, using a smaller gap value of 14.5 meV, higher electron temperature (more photoelectrons are needed to absorb the laser energy) is produced with the same laser fluence comparing to the single crystal. A borderline value of $T_e^*$=19.6 K is obtained at *F*=100 nJ/cm$^2$, which is consistent with the less significant decay profile change observed in thin film. The interesting nonequilibrium situation produced by laser pulses, where electron and lattice temperatures deviate from each other, allow us to assess the nature of the critical temperature $T_2$. At *F*=100 nJ/cm$^2$, $T_p^*$(*T*=5 K) exceeds $T_2$ by far without suppressing the relaxation acceleration, indicating a less significant role of $T_p^*$ in the observation of in-gap states. Thus, the critical temperature $T_2$ observed is likely more strongly affected by electron correlation effects comparing to simple thermal activation. We should point out that chemical potential is the only flexible parameter in our calculation. A Fermi level lying 7 meV below the conduction band is assumed, which provides best match to the measurement results. Variations of chemical potential will not alter the general trend discussed above, but are expected to affect the threshold temperature and fluence required for the surface state observations. Therefore the values of $T_2$ and $F_c$ are likely sample dependent. Based on this model, impulsively injected hot electrons can "turn off" the in-gaps states. This finding suggests the possibility of modulating the low temperature ground state of SmB$_6$ in ultrahigh speed by femtosecond laser.

Here we discuss the possible nature of the observed in-gap states. Besides of topological surface states, chemical potential shift at polar surface can also lead to metallic in-gap states [32]. As pointed out by Frantzeskakis, *et. al.* [33], the two mechanisms can be difficult to distinguish, for example, in ARPES measurements due to the resolution limit. However, we argue that the in-gap states observed here cannot be due to the second mechanism. First, the polar surface states are often temperature independent, though the in-gap states we observe form only below 20 K. Second, higher chemical potential favors the filling of polar surface states, in contrast the in-gap states we observe are destroyed at higher electron temperatures (*i.e.* higher effective chemical potential). Third, while the polar surface states forms pure electrostatically, the non-equilibrium features of the in-gap states we observe have a clear connection to correlation effects. Another remark is that, although analysis considering only uniform bulk properties well explains the data, such consistency doesn't preclude the possible surface nature of the in-gap states we observed. Penetration depth of 400nm pump laser in SmB$_6$ is $d = \frac{\lambda}{4\pi k} \approx 15$ nm (extinction coefficient $k = 2$ [50]). Based on the quantum oscillations observed by torque magnetometry in SmB$_6$ single crystal [27], surface states Fermi velocities $v_F \sim (2.9 \pm 0.4) \times 10^5$ m/s near $\bar{\Gamma}$ point and $v_F \sim (6.5 \pm 0.21) \times 10^5$ m/s near $\bar{X}$ point can be derived [11]. Conversion to topological surface state penetration depth $\xi = \frac{\hbar v_F}{E_g}$ [62] yields values of 9.1 nm and 20 nm, comparable to the laser penetration. Therefore, ultrafast pump-probe measurements can be considered surface sensitive tools with regard to the investigations of surface states in SmB$_6$.

**Conclusion**

By studying the transient carrier dynamics in SmB$_6$, a coherent picture of the band structure evolvement in SmB$_6$ is successfully derived. We found two characteristic temperatures $T_1$ and $T_2$. $T_1$=100 K, corresponding to the opening of *f-d* hybridization gap, is pin-pointed by the abrupt disappearance of *f*-band plasmon. $T_2$=20 K indicates the temperature below which in-gap states form and facilitate faster photocarrier relaxation. The observed in-gap states are strongly affected by electron correlation effects, which exclude the trivial explanation by polar surface layer. Measurement results obtained in single crystal and thin film samples are highly consistent. The hybridization gap and low temperature in-gap states, while deep-rooted in strong correlation physics, are robust against the presence of defects and grain boundaries.

More studies are necessary to conclude whether the observed in-gap states are indeed topological surface states. For example, it will be very interesting to look for plasmon modes associated with these in-gap states. The plasmon dispersion relation and damping properties can provide highly valuable insight into the in-gap states' dispersion and their correlations with the bulk. Also, time resolved measurements of the spin dynamics can also be performed to evaluate the spin-orbit coupling effect and other spin related features predicted by TKI theory.

**Acknowledgements**

The work at West Virginia University is supported by the Department of Energy Grant No. DE-SC-0010399. The work at UC Irvine is supported by National Science Foundation grant DMR-1350122. The work at University of Wisconsin is supported by the National Science Foundation under DMREF Grant No. DMR-1234096.

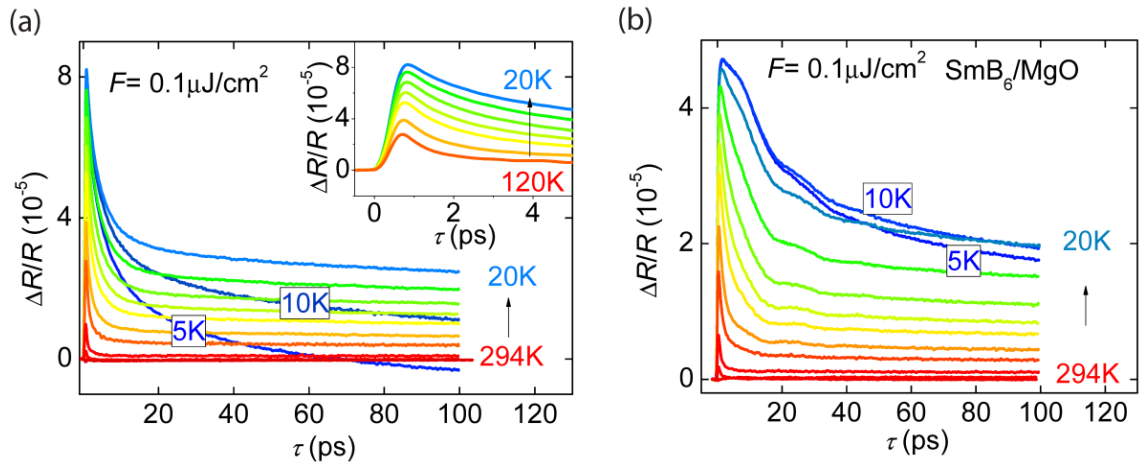

**Figure 1 Temperature dependence of the time-resolved PI reflectivity measured in SmB$_6$** (a) bulk single crystal and (b) thin film. A low laser fluence 0.1 μJ cm$^{-2}$ is used in the measurements.

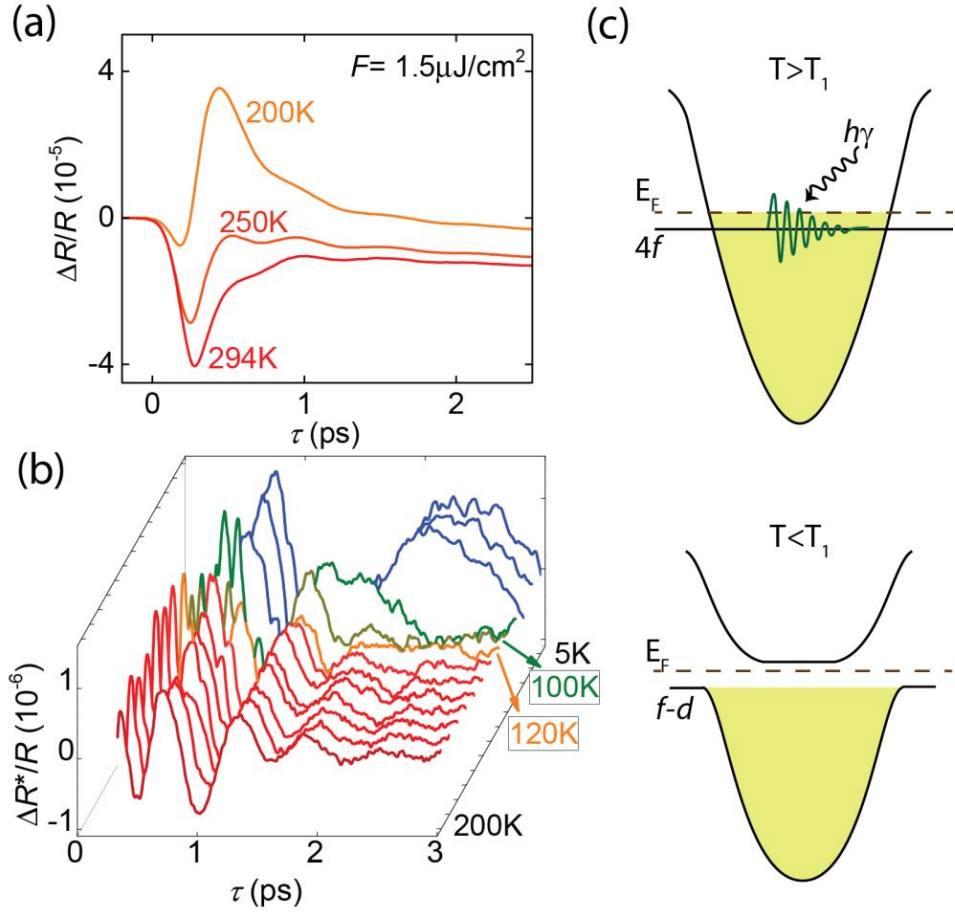

**Figure 2**: **Temperature dependence of *f*-band plasmon mode** (a) PI reflectivity measurement of single crystal at high temperatures, showing a 1.7 THz oscillation signal ($S_1$). (b) Background subtraction is performed to emphasize the temperature dependence of the oscillating features. (c) Illustration of evolvement of $S_1$ at hybridization gap opening.

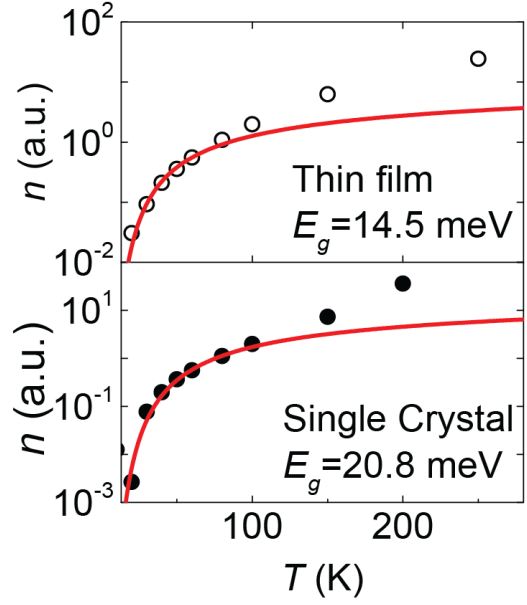

**Figure 3 Temperature dependence of the thermal carrier densities *n(T)*** extracted from the maximum amplitudes of PI reflectivity signals. Red curve is the fitting of: $n(T) \sim T^{1/2} \exp(-\frac{E_g^*}{2kT})$, with the effective gap energies ($E_g^*$) of 20.8 mV and 14.5 meV for single crystal and thin film sample respectively.

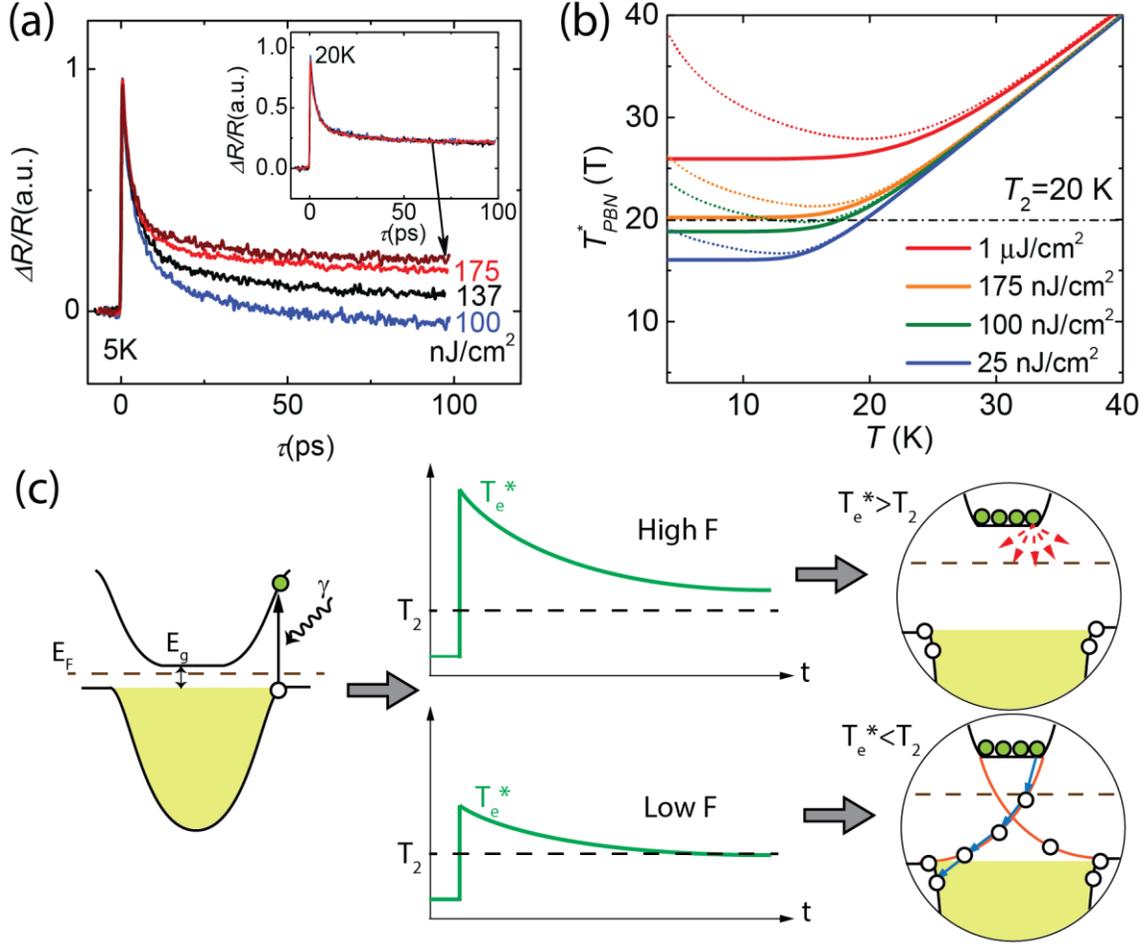

**Figure 4**: **Pump probe signatures of the development of in-gap states at low temperatures**. (a) Power dependence of PI reflectivity measured at 5 K and 20 K (inset). To emphasize changes in the relaxation profile, signal maximum near zero delay is normalized to one. (b) Assuming the presence of phonon bottleneck, effective electron temperature (solid line) and phonon temperature (dotted line) calculated at the quasi-equilibrium between electrons and above-bandgap phonons. Black dash line indicates the critical temperature for in-gap states development ($T_2$=20 K). (c) Illustration of electron relaxation at low temperatures at high and low laser fluences. Right after the initial excitation when the electron temperature is very high, density of states within band gap is zero and phonon bottleneck effect dominates. Depending on the laser fluence, electron temperature ($T_e^*$) at quasi-equilibrium with high frequency phonons varies. When $T_e^*$ is larger than the critical temperature $T_2$, bottleneck resumes. When $T_e^*$ is smaller than $T_2$, in-gap states start to develop and as a result accelerate the electron relaxation.

# Ultrafast observation of electron hybridization and in-gap states formation in Kondo insulator SmB$_6$: Supplementary Information

Sanjay Adhikari[1], Yanjun Ma[3], Zachary Fisk[2], Jing Xia[2], Chang-Beom Eom[3], Cheng Cen[1*]

[1]Department of Physics and Astronomy, West Virginia University, Morgantown, West Virginia 26506, USA

[2] Department of Physics and Astronomy, University of California, Irvine, California 92697-4575, USA

[3]Department of Material Science and Engineering, University of Wisconsin-Madison, Madison, Wisconsin 53706, USA

*Correspondence to chcen@mail.wvu.edu

## Sample synthesis

SmB$_6$ single crystals were grown using the aluminum flux method. These crystals are then inspected using X-ray analysis to make sure SmB$_6$ is the only content. To further ensure that there is no remaining aluminum inside the crystal, we perform susceptibility measurements to check any sign of superconductivity of aluminum. The surfaces of these crystals were carefully etched using hydrochloric acid (50 HCL + 50 DI water) for 2 minutes and then cleaned using solvents to remove the possible oxide layer. Samples used in the experiments have naturally existing well defined (100) surfaces that are a few mm in size. We have carried out transport measurements of crystals grown from the same batch and make sure there is low temperature resistance saturation, a sign of surface-dominated conduction.

Thin film of SmB$_6$ is synthesized by pulsed laser deposition (PLD). A KrF excimer laser outputs 248 nm laser pulses, which is focused on a SmB$_6$ ceramic target with the spot size of 2.8 cm$^2$ in the deposition chamber. The laser repetition rate is 10 Hz, and the energy density is 3.6 J/cm$^2$. The ablated SmB$_6$ materials deposit on a MgO (100) single crystal substrate, of which the surface is polished and the temperature is kept at 850 C. The working distance is 6 cm. The growth pressure is maintained at $1 \times 10^{-7}$ Torr during the deposition. Right after growth, X-ray diffraction on the film reveals its polycrystalline structure; however, it is also noticed that the film changes to amorphous phase over time even when it's stored in a desiccator inside which the temperature is set to be about 20 ℃ and the humidity is controlled around 10%.

**Transport Measurements:**

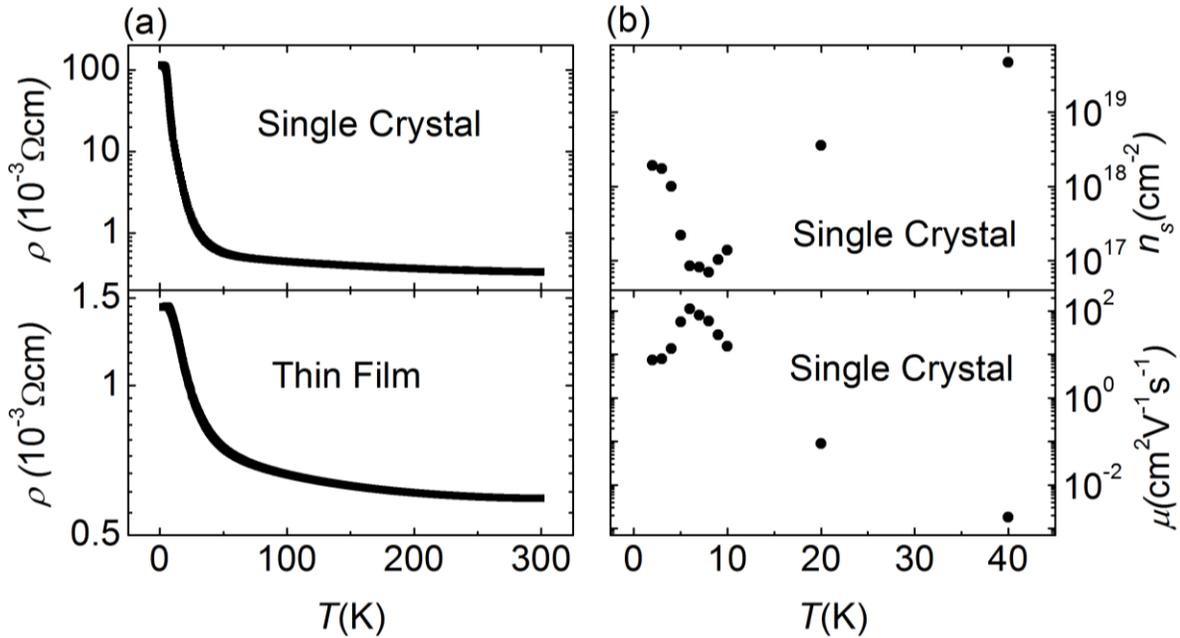

**Figure S1**: Resistivity vs temperature for both (a) single crystal and thin film. Resistivity measurement for the single crystal was done in hall bar configuration and thin film was done in van der pauw configuration. The resistivity saturates at temperatures below 4K, consistent with the existence of surface conducting states. (b) Sheet carrier density and mobility as a function of temperature for the single crystal extracted from the hall voltage measurement.

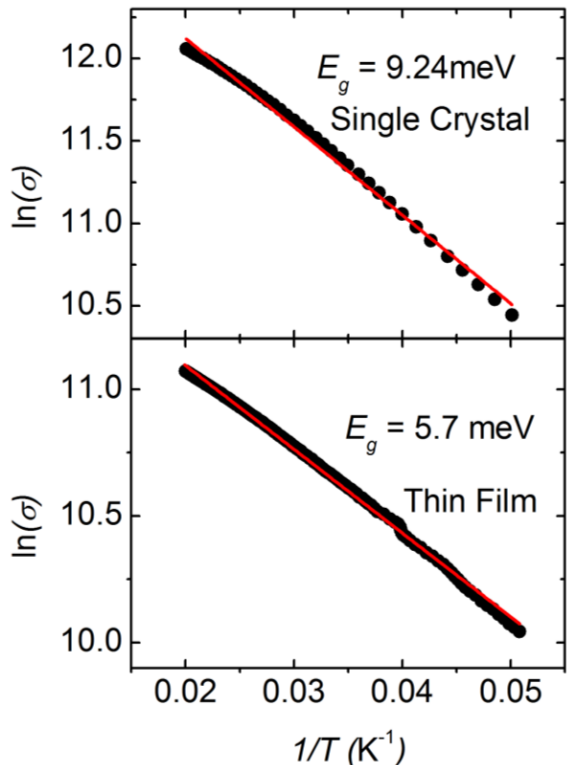

**Figure S2**: ln(σ) vs. 1/T, 20K<T<50K, for single crystal (on the top) and thin film (on the bottom), where σ is conductivity. The fit, shown in red, is done using thermal activation law equation given by $\sigma(T) = \sigma_0 \exp(-E_g / 2kT)$. Where, k is the Boltzmann constant, and $E_g$ is the energy gap.

**Raman Measurements**

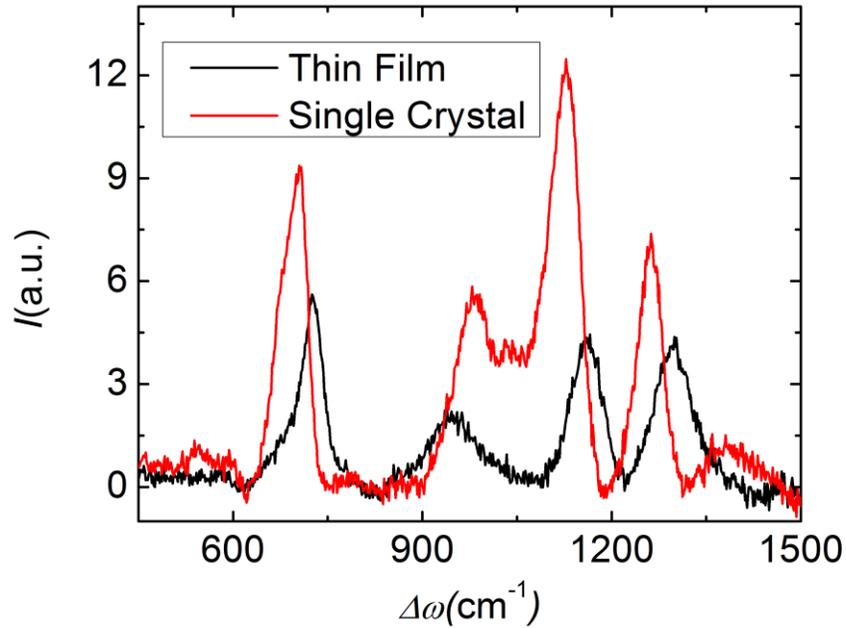

**Figure S3**: Room temperature Raman spectra from thin film and the single crystal. Both thin film and single crystal show the three main Raman active phonon modes of $A_{1g}$ (1263 cm$^{-1}$), $E_g$ (1128 cm$^{-1}$) and $T_{2g}$ (706 cm$^{-1}$) symmetry (Nyhus, Copper, Fisk, & Sarrao, 1997). These modes are associated with the displacement of boron atoms in $SmB_6$. There is one major unknown mode at 979 cm$^{-1}$. This mode, present in both thin film and single crystal, could be arising from residual photoresist after photolithography process (Potma et, al., 2004). All of the major Raman active modes in thin film sample are shifted, with less than 10%, to lower wavenumber. This could be due to the possible stress in the thin film and other crystalline quality related reasons.

**Strain Wave**

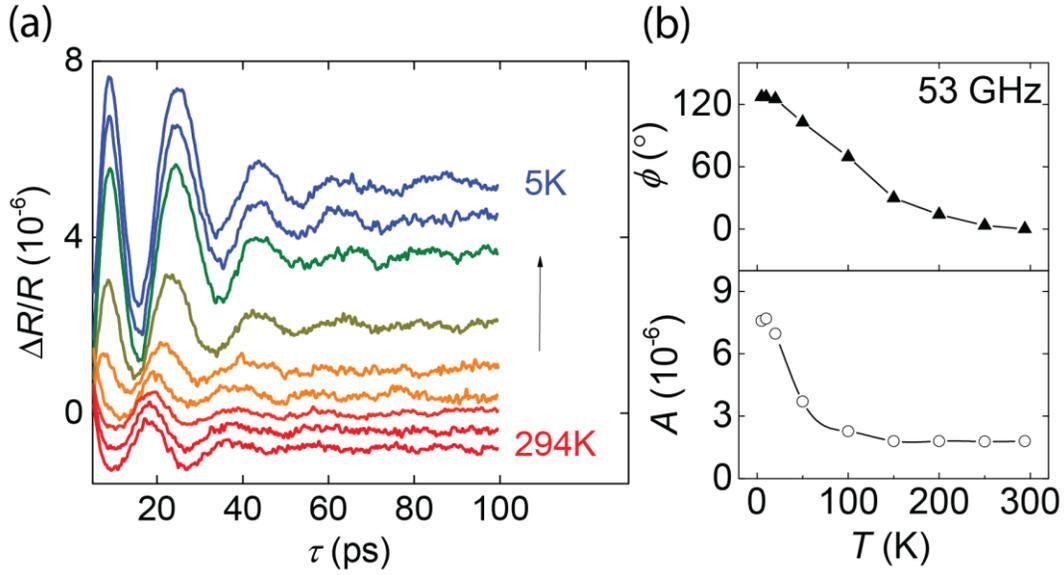

**Figure S4**: Strain wave in thin film sample (a) Temperature dependence of the 53 GHz strain wave oscillations extracted from the PI reflectivity measurements. (b) Amplitude of the oscillations sharply increases at 100 K and the phase gradually varies with the change of temperature. Pump fluence of 1.5μJ/cm$^2$ is used in the measurements.

A damped oscillation, $S_2$, of much lower frequency (53GHz) is uniquely present in the thin film (Fig.1(b), Fig. S4(a)). We attribute $S_2$ to the optically excited strain wave which reflects repeatedly at the top and bottom surface of the film (Thomsen, Grahn, Maris, & Tauc, 1986). Surface absorption of the pump pulse induces a local thermal stress ($\sigma$) and initializes the strain wave propagating and reflecting inside the film. The resultant periodically changing strain at the film surface transiently induces changes in the local density of states (LDOS) and surface deformation, both resulting in the modulation of measured probe pulse reflection. Considering a model where the top surface of the film is free and the SmB$_6$/MgO interface is fixed by the substrate, the expected oscillation frequency is $f = v/4d$ (Thomsen, Strait, Vardeny, Maris, Tauc, & Hauser, 1984), where $v$ is the speed of sound and $d$ is the film thickness (65nm). According to this formula, $f = 53$GHz corresponds to a sound speed of $1.4\times10^4$ m/s in SmB$_6$, which is high but reasonable given the well-known hardness of borides. While the frequency remains constant throughout the measurements, phase shift of $S_2$ is observed upon cooling, which is owing to the variance of ratio between stress induced changes in refractive index and extinction coefficient (Thomsen et al., 1986). More interestingly, amplitude of $S_2$ experiences a sharp upturn at 100 K. Based on a simple isothermal model, pump induced thermal stress can be expressed as $\sigma = -3(Q/C)\beta K$, where $Q$ is the energy deposited by a single pump pulse, $C$ is specific heat, $\beta$ is thermal expansion coefficient and $K$ is the bulk modulus. Since there is no evidence of discontinuous change of the involved parameters around 100 K (Mandrus et al., 1994; Phelan et al., 2014). The upturn in $S_2$ amplitude is more likely due to the enhanced effect of stress on the local density of states at low temperatures, and possibly is related to the gap opening at 100 K. It has been reported that uniaxial stress (application of pressure) can effectively modulate the hybridization gap size (Cooley, Aronson, Fisk, & Canfield et al., 1995; Gabáni et al., 2003; Derr et al., 2008). With $Q = 1\times10^6$ Jm$^{-3}$ and literature reported values of $C = 4\times10^5$ JK$^{-1}$m$^{-3}$ (Phelan et al., 2014), $\beta = -5 \times 10^{-6}$ K$^{-1}$ (Mandrus et al., 1994), $K = 147$ GPa (Sandeep et al., 2012), a rough estimation yields $\sigma \sim 0.1$ kbar at low temperatures, which is significant considering that a pressure larger than 40 kbar can suppress the gap completely(Cooley et al., 1995; Gabáni et al., 2003). We note that the thermal stress not only generate changes in the LDOS close to Fermi levels but may also impact the higher energy states, and the overall effects contribute to the change in 800nm probe's reflectivity approaching $10^{-5}$ level at a pump fluence of 1.5μJ/cm².

**Laser heating simulation – Equilibrium Temperature Increase**

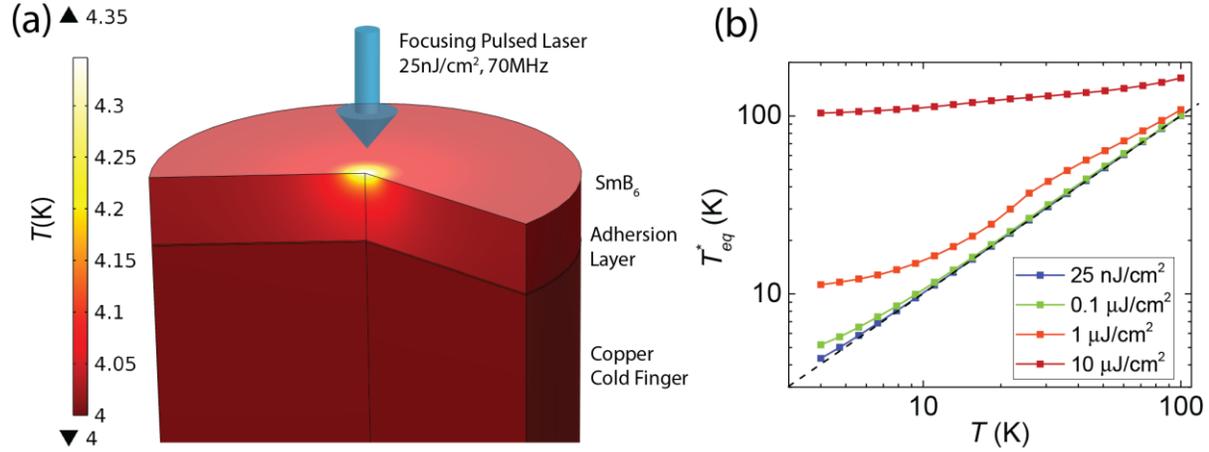

**Figure S5**: Finite element simulation of laser heating effect. (a) Laser illuminated sample temperature distribution at thermal equilibrium with a cold finger temperature of 4 K. (b) Plots of the simulated effective thermal equilibrium temperatures at the center of laser focus ($T_{eq}^*$) as a function of actual temperature measured at the cold finger (*T*). Curves with different colors correspond to different laser fluences. As reference, dash line indicates the effective temperature without any external heating ($T_{eq}^* = T$). Significant laser heating at low temperature is expected when laser fluence exceeds 1 μJ/cm².

We model $T_{eq}^*$ in our cryostat setup by solving thermal conduction equation using finite element method. A basic thermal transfer model of $\rho C_p (\frac{\partial T}{\partial t} + u \cdot \nabla T) = \nabla \cdot (\kappa \nabla T) + Q$ is considered, where $\rho$ is the density, $C_p$ is the thermal capacity, *T* is the temperature, *t* is the time, *u* is the velocity vector, $\kappa$ is the thermal conductivity, and *Q* is the external heat source. 400 nm laser pulses with 70 MHz repetition rate is focused at the top of sample surface. Laser intensity distribution at the focal spot is assumed to be Gaussian with a full width at half maximum (FWHM) of 200 μm. A reflectivity of 30% is used for 400nm laser. Literature reported temperature dependent thermal capacity (Phelan et.al., 2014) and thermal conductivity (Popov, Novokov, Sidorov, & Maksimenko, 2007) data of SmB$_6$ single crystal is used in the simulation. A 10 μm adhesion layer of VGE-7031 is also included in the model. Its technical data is taken from the vendor's website (http://www.lakeshore.com/products/cryogenic-accessories/varnish/pages/Specifications.aspx).

**Laser heating simulation – Transient Temperatures Increase**

Following the effective temperature method developed by Parker (Parker 1975), we calculate the effective electron temperature $T_e^*$ and effective phonon temperature $T_p^*$ at the quasi-equilibrium between electrons and above-bandgap phonons (phonon bottleneck). Under the framework of RT model, we assume that the energy deposited by laser pulses is stored locally in electrons and above bandgap phonons (ignoring slow processes of diffusion and photon emissions):

$$\Delta E = \Delta E_e + \Delta E_p \qquad (1)$$

Effective transient temperatures $T_e^*$ and $T_p^*$ are defined as following:

$$\Delta E_e = \int_0^\infty (2\epsilon + E_g) D(\epsilon)(f(\epsilon + \mu, T_e^*) - f(\epsilon + \mu, T))d\epsilon,$$
$$\Delta E_p = \int_{E_g/\hbar}^{\omega_D} \frac{3\omega^2}{2\pi^2 v^3} \hbar\omega (F(\omega, T_p^*) - F(\omega, T))d\omega \qquad (2)$$

where $D(\epsilon)$ is the electron density of states, $f(\epsilon, T)$ and $F(\omega, T)$ are Fermi-Dirac distribution for electrons and Bose-Einstein distribution for phonons, $\mu$ is the chemical potential, $v$ is the speed of sound, and $\omega_D$ is the Debye frequency derived from $v$. Following similar convention, electron and above band gap phonon densities can be expressed as:

$$N_e(T) = \int_0^\infty D(\epsilon) f(\epsilon, T) d\epsilon,$$
$$N_p(T) = \int_{E_g/\hbar}^{\omega_D} \frac{3\omega^2}{2\pi^2 v^3} F(\omega, T) d\omega \qquad (3)$$

The detailed balance between electrons and above bandgap phonons are determined by the coupled differentiation equations:

$$\frac{dN_e}{dt} = \eta N_p - R N_e^2$$
$$\frac{dN_p}{dt} = -\eta N_p/2 + R N_e^2/2 - \gamma(N_p - N_p^T) \qquad (4)$$

where $N_p^T$ is the phonon density at thermal equilibrium, $\eta$ is the rate of electron-hole pairs creation from above bandgap phonons, $R$ is the rate of phonon emission from electron-hole pair recombinations, and $\gamma$ indicates the phonon decay rate by anharmonic processes or diffusion. When $\eta \gg \gamma$, this lead to

$$\left(N_e(T_e^*)/N_e(T)\right)^2 = N_p(T_p^*)/N_p(T) \qquad (5)$$

Values of $T_e^*$ and $T_p^*$ at different base temperatures and laser fluences are obtained by solving equation (1-3,5) numerically and self consistently. In our calculations, we take a literature reported conduction band effective mass of $100m_0$ (Gorshunov, et. al., 1999). The values of gap energy and speed of sound are obtained from the previously discussed RT model fitting and strain wave analysis. The only flexible parameter in the calculation is the chemical potential.